\documentclass[aps,preprint,prl,superscriptaddress,longbibliography]{revtex4-1}
\usepackage[usenames,dvipsnames]{color}
\usepackage{graphicx,textcase}
\usepackage{dcolumn,overpic}

\begin{document}

\title{Electronic topological transition and non-collinear magnetism in compressed hcp Co}
\author{Y.O. Kvashnin, W. Sun, I. Di Marco, and O. Eriksson}
\affiliation{ Department of Physics and Astronomy, Division of Materials Theory, Uppsala University, Box 516, SE-75120 Uppsala, Sweden}

\begin{abstract}
Recent experiments showed that Co undergoes a phase transition from ferromagnetic hcp phase to non-magnetic fcc one around 100 GPa.
Since the transition is of first order, a certain region of co-existence of the two phases is present.
By means of \textit{ab initio} calculations, we found that the hcp phase itself undergoes a series of electronic topological transitions (ETTs), which affects both elastic and magnetic properties of the material. 
Most importantly, we propose that the sequence of ETTs lead to the stabilisation of a non-collinear spin arrangement in highly compressed hcp Co. 
Details of this non-collinear magnetic state and the interatomic exchange parameters that are connected to it, are presented here.
\end{abstract}

\maketitle

By the development of the diamond anvil cells, it became possible to carry out high pressure experiments upon to a few hundreds GPa.\cite{Boehler-HP} 
Such compressions roughly correspond to the state of the matter close to the Earth's core, so that the processes occurring under the deep mantle can be reconstructed in nowadays laboratories.\cite{Dubrovinsky-FeNi-2007,Ni-stable-150GPa} 
Particularly, the investigation of magnetic systems is of interest, because it can give fundamental insights on the origins of the geomagnetism.\cite{geodynamo}

Cobalt is not a particularly abundant element in the Earth's core, and is therefore not intensively studied under extreme conditions. 
However, it is an outstanding element from the technological and scientific viewpoints. 
At ambient conditions, Co crystallizes in an hcp lattice showing a ferromagnetic (FM) order with a high Curie temperature ($T_c$) of 1388 K. 
Its Fermi surface (FS) is highly spin-polarised, which makes it a great spin-filter material. The hcp phase can be stable up to a very high pressure, but at around 100 GPa an hcp to fcc transition takes place.\cite{Iota-hcpCo-210GPa,Ishimatsu-paramag-Co}
There is an on-going debate about the magnetic state of cobalt above this critical pressure.
Iota \textit{et al.} proposed that Co remains in the magnetic hcp phase at least up to 100 GPa. 
For higher pressures, hcp Co is gradually transformed into the non-magnetic (NM) fcc phase, which identifies a region of 50 GPa where the two phases coexist and Co gradually loses its magnetism.\cite{iota-3dMe-GPa} 
On the other hand, Torchio \textit{et al.} argued that the magnetism is already completely lost at 120 GPa in the mixed phase.\cite{Torchio-Co-120GPa} This is also in qualitative agreement with more recent results reported by Ishimatsu \textit{et al.}, but these authors also suggest the existence of a super paramagnetic (PM) fcc phase in the high-pressure region above 135 GPa.\cite{Ishimatsu-paramag-Co}
All aforementioned groups analysed the x-ray magnetic circular dichroism (XMCD) data, measured at the K edge. 
It is worth mentioning that this type of experiments does not probe directly the spin moment, but rather the weak moment of the $4p$ states.\cite{Torchio-ni-prl} 
Thus, some small values of the total magnetisation might appear as a negligible noise on the spectra.

Theoretically, the problem of magnetism of Co under pressure has been extensively investigated by means of the density functional theory (DFT).\cite{Yoo-dhcp,Steinle-prb99,DFT-hcp-fcc-Co,DFT-Co-phonons,Steinle-hcpCo,Kuang} 
Nevertheless, the definition of the transition pressure is still uncertain, which can be traced back to differences among the employed DFT implementations.\footnote{The results of the Ref.~\onlinecite{Steinle-hcpCo} indicate that the hcp phase loses its magnetism before the hcp-fcc transition, which is in contradiction to the conclusions from all other DFT-based studies.} 
The most recent calculations indicate that the fcc phase favours a NM state before the transition, while the hcp phase remains magnetic up to more than 160 GPa.\cite{Kong-hcpCo} 
The FM-NM transition in the hcp phase was identified to be of second order, while in fcc phase it is a first-order one.\cite{DFT-hcp-fcc-Co}

On the experimental side, the pressure-induced anomalies of the $c/a$ ratio\cite{antonangeli-Co-APL2008} and Raman frequencies\cite{Goncharov-hcpCo-phonons} were attributed to a strong magneto-elastic coupling, which is argued to be an intrinsic feature of the material. 
However, any unequivocal evidence for the coupling between magnons and phonons have not been presented yet. 
In this Letter, we point out that the above-mentioned peculiarities result from an electronic topological transition (ETT), or the so-called Lifshitz $2\frac{1}{2}$ transition.\cite{Lifshitz-ETT} 
We also argue that the change of the FS gives rise to a non-collinear magnetic ground state in hcp Co under pressure, and that this may be an excellent way to experimentally detect the influence of the ETT.

In the present study we have used several DFT implementations.
The structural optimisation for both hcp and fcc phases was performed using the Vienna \textit{Ab initio} Simulation Package (\texttt{VASP}).\cite{VASP,vasp-paw} 
For this purpose, the stress tensor was analysed to reach a certain value of external pressure. 
The lattice parameters were converged by using a large plane-wave energy cut-off of 600 eV and very dense \textbf{k}-point grids,
i.e. 45$\times$45$\times$27 and 45$\times$45$\times$45 for respectively hcp and fcc structures.
Then, the obtained crystal structures were used to perform DFT calculations by means of the 
full-potential linear muffin-tin orbital method (FP-LMTO) as implemented in the $\texttt{RSPt}$ code\cite{rspt-book}.
These simulations were used to extract the effective exchange parameters ($J_{ij}$) and calculate the corresponding adiabatic magnon spectra,
 as well as for analysing the FS. The $J_{ij}$'s were computed between the $3d$ states, 
which were projected onto the muffin-tin spheres (for details see Ref.~\onlinecite{jijs-rspt}).
Additional simulations of the spin spirals (SS) were performed via the \texttt{PY-LMTO} code.\cite{py-lmto} 
All results were obtained with the GGA-PBE\cite{gga-pbe} exchange correlation functional. 
Within this framework, the experimental hcp-fcc transition pressure was shown to be well reproduced.\cite{DFT-hcp-fcc-Co,DFT-Co-phonons}
We did not consider the effects of strong correlations, since it would require to know not only the value of the Hubbard $U$ parameter, but also its pressure dependence.
Moreover, in our study we have to deal with very high pressures, where the large band broadening should diminish the correlation effects.
Spin-orbit coupling is neglected throughout the paper, however we discuss how it possibly affects the results.

\begin{figure}[!h]  
\includegraphics[angle=0,width=60mm]{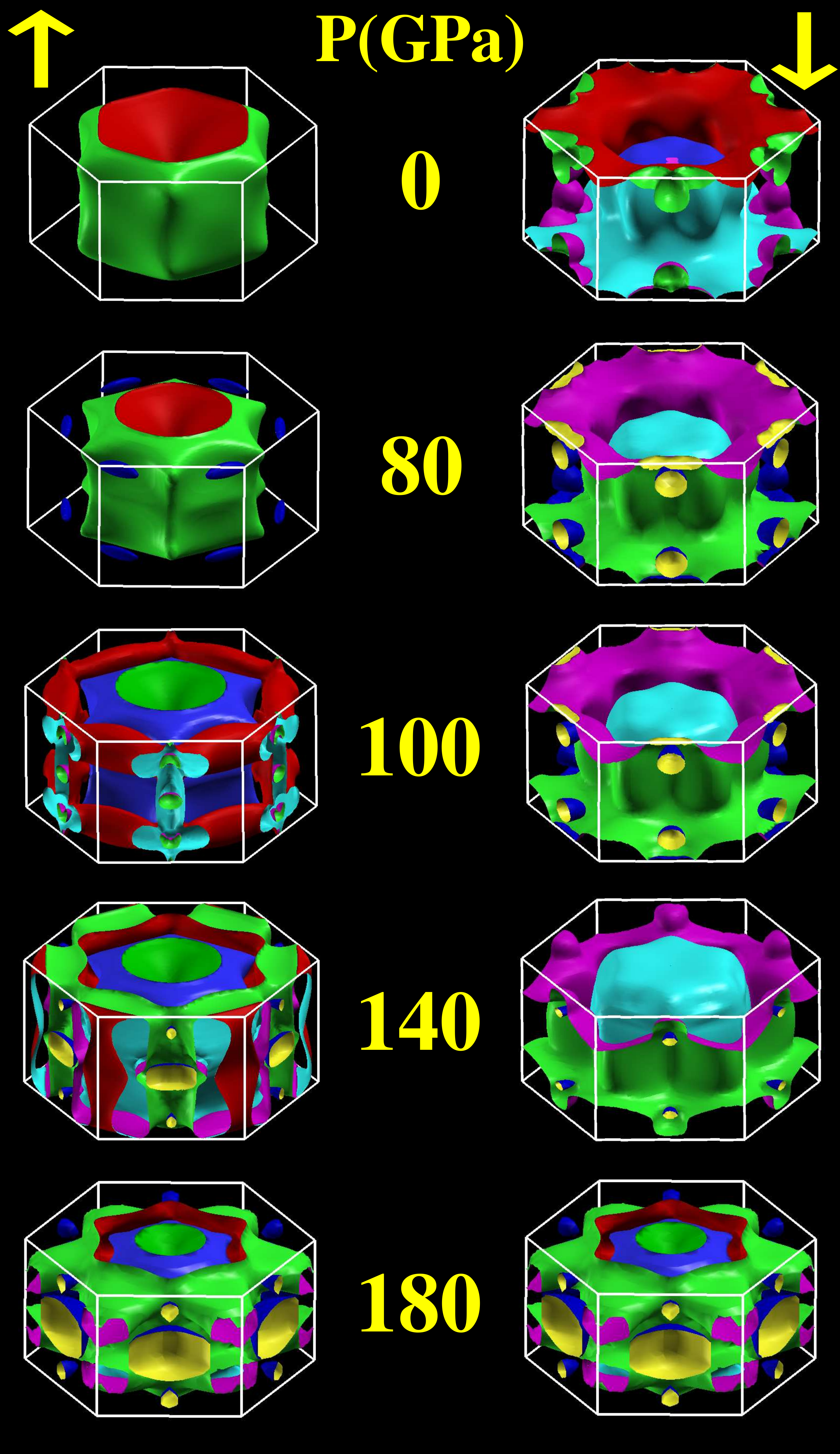}
 \caption{(Color on-line) Fermi surface of the majority-spin ($\uparrow$) and minority-spin ($\downarrow$) electrons in hcp Co for different values of external pressure (in GPa). The results were obtained with the GGA-PBE functional. The plot was produced with the \texttt{XCRYSDEN}\cite{xcrysden} software.}
\label{fermisurf}
\end{figure}

The calculated FS of the ferromagnetic hcp Co as a function of pressure is shown in Fig.~\ref{fermisurf}.
At ambient conditions, four minority bands and two majority-spin ones cross the Fermi level ($E_F$). 
The shape of the FS is barely changed up until roughly 80 GPa (see Supplemental Material (SM)\footnote{See Supplemental Material at [URL will be inserted by publisher] for the additional results of the calculations} for more details).
Above 80 GPa, a sequence of ETTs happen at different pressures in the spin-up channel: at around 80, 100 and 140 GPa, one more band is counted to contribute to the $E_F$. 
At 180 GPa the NM state is reached, which is characterised by all five bands in each spin channel crossing $E_F$.

Close to 80 GPa a particularly important ETT takes place.
Many small sheets emerge in the spin-up channel, and the part of the FS close to the $\Gamma$ point of the spin-down channel is also significantly modified. 
This causes a transition from a strong ferromagnet to a weak one. 
Although there is no direct experimental evidence of this ETT, the changes in the topology of the FS can explain the changes that have been observed in several other quantities at around 80 GPa. 
First of all, at around this pressure, the $c/a$ ratio alters its pressure derivative, as was unveiled by Antonangeli \textit{et al},\cite{antonangeli-Co-APL2008} who confirmed it both experimentally and theoretically.
Second, vibrational properties, such as e.g. sound velocities, show anomalous behaviour at around 75 GPa.\cite{antonangeli-Co-PRB05} 
This anomaly is accompanied by a change of the slope of the $E_{2g}$ phonon modes, which was first measured by Goncharov \textit{et al.} (Ref.~\onlinecite{Goncharov-hcpCo-phonons}) and later on reproduced in the DFT calculations.\cite{DFT-Co-phonons}
Moreover, previous DFT studies\cite{antonangeli-Co-APL2008,Kong-hcpCo,Steinle-hcpCo} suggested that at the same compression ratio, i.e. $V/V_0\approx$0.8, the pressure-driven reduction of the spin moment ($M_S$) becomes faster. These anomalies in the $c/a$ ratio and $M_S$ are very well reproduced in our calculations, as reported in the SM.\cite{Note2} 
Most importantly, in combination with our results on the FS, they can offer a strong evidence for the existence of the ETT, and can be seen as signatures of this 2$\frac{1}{2}$ order transition.
At low pressure, hcp Co has a majority-spin band completely occupied, which results in a weak pressure dependence of $M_S$.
However, the ETT pushes up the majority-spin band towards $E_F$, so that $M_S$ becomes more vulnerable to external stimuli, e.g. to pressure (for a detailed explanation of this mechanism, see Ref.~\onlinecite{P.Mohn-magnetism-book}). 
The resulting decrease of the $M_S$ is the primary reason for the destabilisation of the FM hcp phase at high pressures.
This is similar to the bcc-hcp transition in Fe, which can be explained by mapping of the DFT solution on a simple Stoner model.\cite{andersen-3dMe-structures}
Once the energy gain due to magnetic moment formation, which is proportional to $(M_S)^2$, gets too small, the system transforms into more close-packed structure.

\begin{figure}[!h]  
\includegraphics[angle=0,width=80mm]{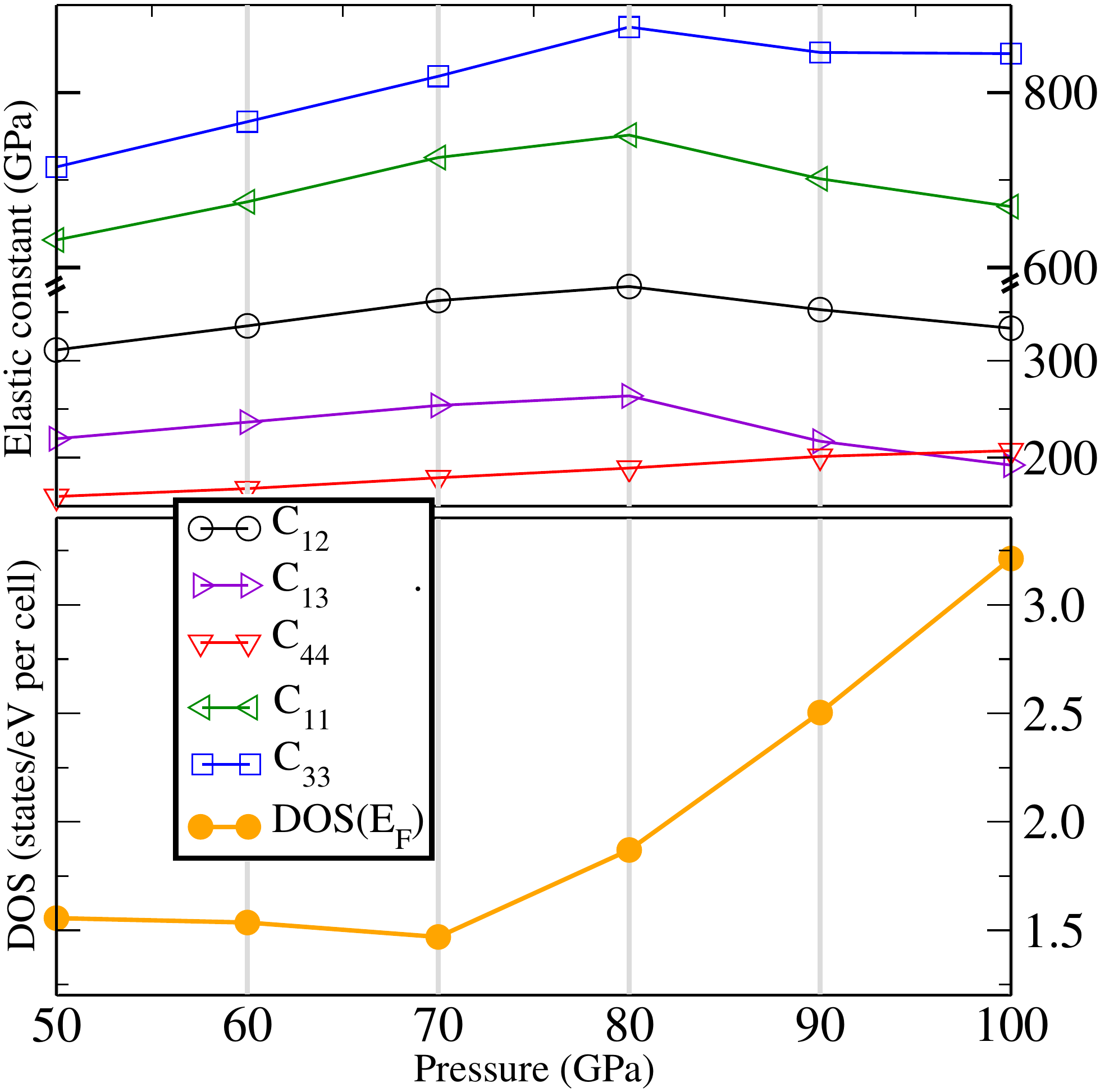}
\caption{(Color on-line) Calculated elastic constants (upper panel) along with the DOS at the $E_F$ (lower panel) as a function of pressure. DOS is summed up for both spin channels.}
\label{ela-prop}
\end{figure}

Signatures of the ETT can also be found in the elastic constants, which are reported in Fig.~\ref{ela-prop}
together with the density of states (DOS) at the $E_F$. 
First of all, one can see that below 70 GPa DOS($E_F$) is weakly dependent on the applied pressure. 
At 80 GPa, a small hump is visible, which is a manifestation of the ETT, as discussed previously. 
After the ETT, the DOS($E_F$) becomes strongly pressure-dependent, and quickly surges upon compression. 
As illustrated in Ref.~\onlinecite{ela-ett}, there is clear link between ETT and lattice properties,
and the elastic constants result to be directly related to the derivative of the DOS at the $E_F$. 
In the case of hcp Co we observe that all elastic constants except $C_{44}$ change their trends around 80 GPa. 
Similar findings have already been reported in a previous study by Kuang \textit{et al.}\cite{Kuang}
But from our results it becomes apparent that this happens as a result of a remarkable change of the DOS($E_F$), associated with the ETT, which appears at the same pressure.

In the literature, these anomalies in the magnetic, elastic and vibrational properties of Co were associated with a magneto-elastic coupling, without any clear explanation.\cite{Goncharov-hcpCo-phonons,Steinle-hcpCo,DFT-Co-phonons,antonangeli-Co-PRB05,antonangeli-Co-APL2008}
In our work we provide a strong evidence that all these changes are consequences of the ETT. 
Present scenario is analogous to the one proposed for PM hcp Fe.\cite{hcpFe-ETT}
In Ref.~\onlinecite{hcpFe-ETT} it was reported that only way to observe the ETT was to augment the DFT solution with the dynamical mean field theory (DMFT)\cite{infdim-DMFT}.
DFT+DMFT\cite{kotliar-DMFT} significantly improves the description of the PM state as compared to the bare DFT, due to its proper account of the local moment fluctuations.
The latter ones give rise to a strong damping of the quasiparticles and the pronounced correlation effects are expected.
Our study, however, concerns the ordered phase where the quasiparticles close to the $E_F$ have much longer lifetimes and the many-body effects are moderate.\cite{grechnev-FeCoNi-PRB}

Note that ETTs do not necessary lead to the anomalies, for instance, in the $c/a$ ratio. 
An example of such situation is hcp Os. 
One of the reasons for this is that the value of the DOS($E_F$) is smoothly dependent on the pressure even when the ETT occurs.\cite{Os-ETT} 
In our case these changes are more conspicuous (see Fig.~\ref{ela-prop}). 
Another reason is probably a weak coupling between the electrons and the lattice. 
On the contrary, it was recently shown that the electron-phonon coupling in hcp Co is five times larger than that in bcc Fe.\cite{hcpCo-el-phon-cplng}
In our opinion, a combination of these two factors in hcp Co allows the ETT to influence the lattice-related properties of the material.

\begin{figure}[!h]  
\includegraphics[angle=0,width=80mm]{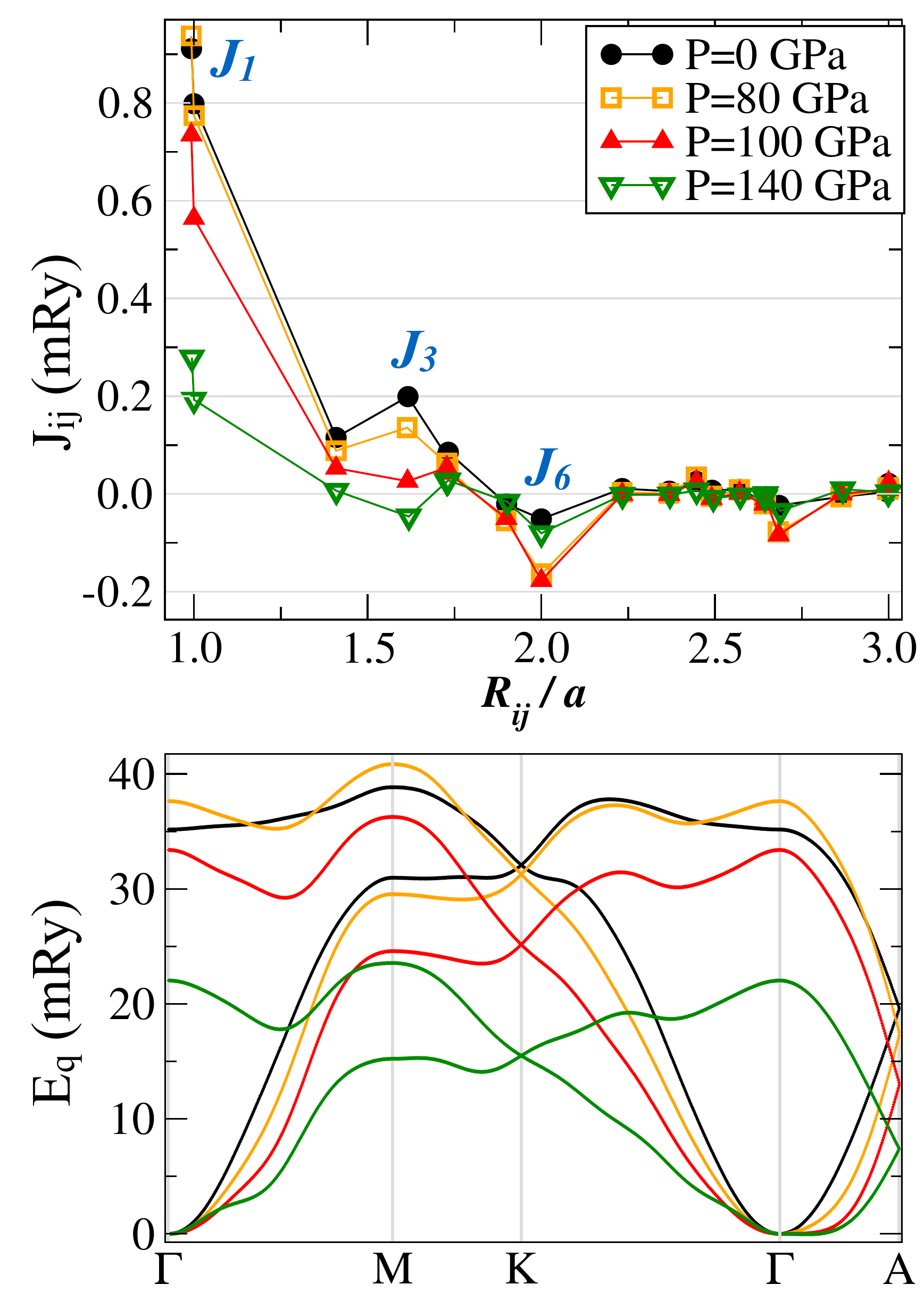}
\caption{(Color on-line) Upper panel: Inter-site exchange parameters in hcp Co as a function of distance for several values of pressure. Lower panel: Calculated adiabatic magnon spectra, obtained using the $J_{ij}$'s shown in the upper panel. The $J_{ij}$'s are extracted for the Heisenberg model with unitary spins (see e.g. Ref.~\onlinecite{jijs-rspt}.)}
\label{hcpCo-jijs}
\end{figure}

In the range of 80-180 GPa the hcp phase of Co remains magnetic and its spin moment strongly depends on the volume (see SM\cite{Note2}).
However, the stability of the FM state at such conditions has never been examined. 
To shed light on this, we have calculated the effective $J_{ij}$-parameters by means of the magnetic force theorem.\cite{lichtenstein-exch}
The exchange parameters within the first few neighbouring shells, calculated at various pressures, are shown in Fig.~\ref{hcpCo-jijs} (upper panel).

The zero-pressure values are in good agreement with prior DFT studies.\cite{Antropov-Co,Turek-Co}
When going from 0 to 80 GPa, one can see that the nearest-neighbour couplings ($J_1$) remain
rather close to each other. This difference is instead most pronounced for the $J_6$ interaction:
it is antiferromagnetic at ambient conditions and gets strongly enhanced at 80 GPa.
This is the primary reason for the emergence of magnon softening along the $\Gamma-A$ direction, as illustrated in the adiabatic magnon spectrum reported in the lower panel of Fig.~\ref{hcpCo-jijs}.
Further compression leads to vast modifications of the exchange parameters. 
Ferromagnetic $J_1$ and $J_3$ couplings become suppressed and the latter one eventually changes sign.
At 140 GPa the $M_S$ is about 0.7 $\mu_B$ per atom, i.e. the system is still magnetic, but its FM state is clearly unstable. 
The global minimum is formed at around $\vec q=(0,0,0.23\pi/c)$, which is along $\Gamma-A$ (or $z$) direction.

\begin{figure}[!h]  
\includegraphics[angle=0,width=80mm]{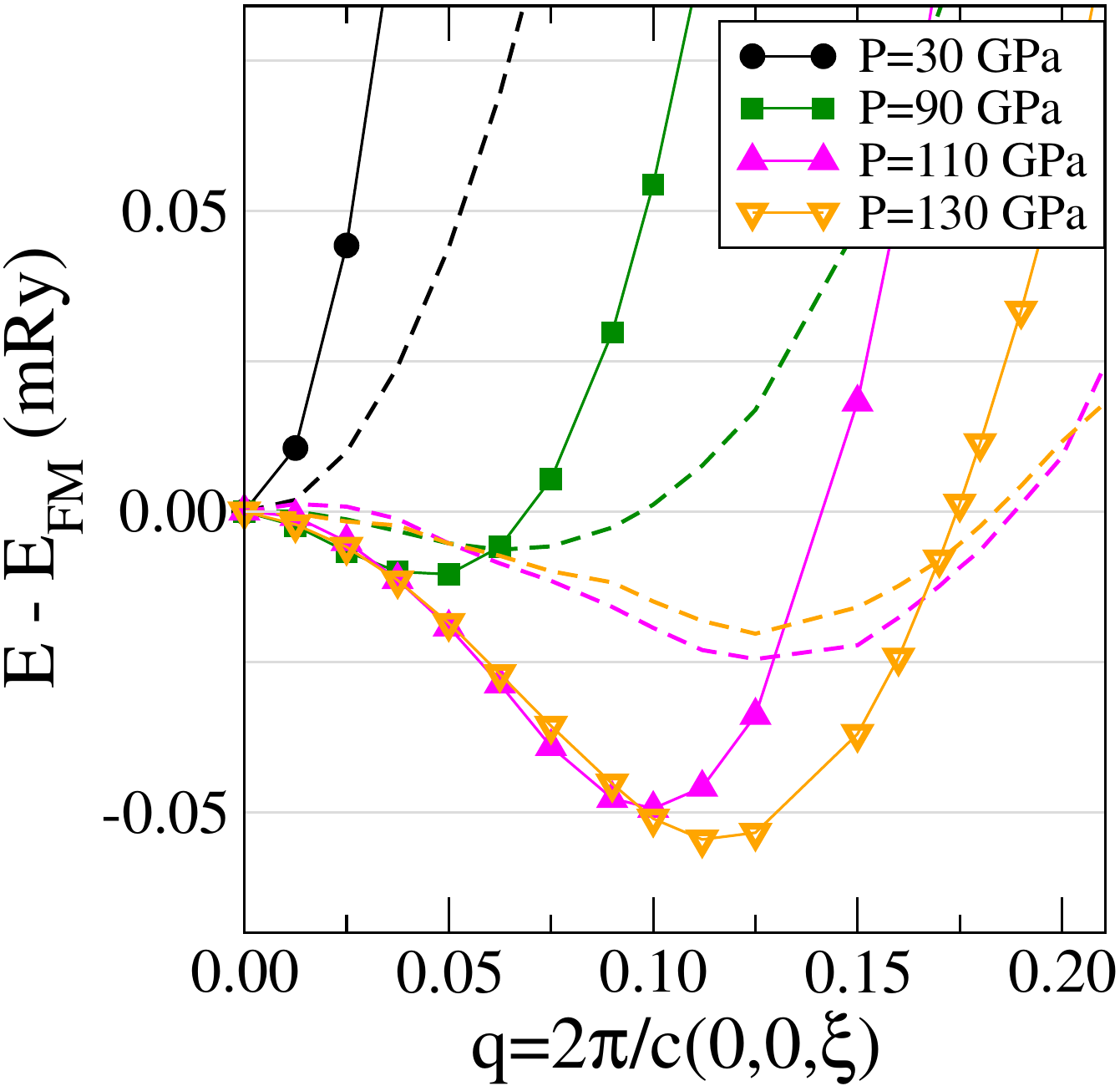}
\caption{(Color on-line) Calculated total energies (relative to that of the FM state) of the spin spiral states propagating along $\Gamma$-A direction. At each pressure the calculations for two different cone angles ($\Theta$) were performed. Dashed lines correspond to $\Theta$=30 degrees, solid ones to $\Theta$=90 degrees. No peculiarities were found in other directions and thus they are not shown.}
\label{spirals}
\end{figure}

The instabilities in the magnon spectra, obtained under a certain pressure, indicate that the long-wavelength excitations destroy the FM ground state. 
What is peculiar is that they have only been found along the $\Gamma-A$ path in the Brillouin zone (BZ), which motivated us to simulate the single-$\vec q$ family of the magnetic states.
Hence, we have calculated the total energies of a set of the SS configurations by means of the generalised Bloch theorem.\cite{Gen-Bloch-theorem}
Any SS state is characterised by two parameters: propagation vector ($\vec q$) and the cone angle between the direction of the propagation and the magnetisation ($\Theta$).
In Fig.~\ref{spirals} we show the relative energies of the SS states with respect to the FM one.

An inspection of Fig.~\ref{spirals} suggests that at low compressions (e.g. 30 GPa) any SS state has a higher energy than that of the FM solution.
This is consistent with the stable spin wave excitation spectra in this regime.
However, this behaviour is changed as the pressure is raised up to 90 GPa. In a certain range of $\vec q$ vectors the SS states become energetically more favourable.
The family of states with $\Theta$=90 degrees seem to be more stable than all other considered states. 
Further increase of the pressure shifts the total energy minimum towards higher $\vec q$-values.
For instance, at 130 GPa the minimum becomes even more pronounced and the total energy difference with respect to the FM state reaches 0.05 mRy.
We have also performed the calculations for the spirals propagating along other directions in the BZ, but no solutions having lower total energy were found. 
Hence the $z$ direction is particularly favourable for the propagation of the SS's. 
The reasons for it can be seen from the analysis of the band structure, presented in SM.\cite{Note2}

Thus we suggest that the non-collinear states should be present in hcp Co upon compression.
In fact, it was previously proposed that it should be possible to stabilise the SS configurations in all ferromagnetic transition metals upon compression.\cite{SS-in-3dMe}
However, this situation is very seldom observed experimentally, due to the presence of the structural transitions and/or the collapse of the magnetic moment at lower pressures.
According to Ref.~\onlinecite{SS-in-3dMe}, the occurrence of the non-collinear spin states is governed solely by the FS topology.
In particular, it is absolutely necessary to have both spin-up and spin-down bands crossing $E_F$.
We believe that hcp Co under pressure is a prototype system, where the ideas, proposed in Ref.~\onlinecite{SS-in-3dMe}, are realised.
As we have shown above, at around 80 GPa one of the spin-up bands start to cross the Fermi level, which leads to the stabilisation of the SS state.
It was recently shown that under certain pressurizing conditions the pure hcp phase of Co can be preserved up to at least 160 GPa.\cite{hcpCo-hydrostat} 
Thus the proposed non-collinear states should exist in a wide pressure range, which in principle leaves enough room for their experimental detection.

Our results indicate that the behaviour of cobalt under pressure is in many ways reminiscent to that of hcp Fe. 
In the PM state of Fe the $c/a$ anomaly was shown to originate from the pressure-driven ETT.\cite{hcpFe-ETT} 
Moreover, the emergence of non-collinear SS states was also suggested from the analysis of the band structure.\cite{hcpFe-noncoll-orig,noncol-hcp-Fe}
Extensive investigation of hcp Fe was partially motivated by the discovery of superconductivity in this system below 1.5~K.\cite{hcpFe-supercond}
A close similarity of the two systems implies that an experimental investigation of the low-temperature properties of hcp Co would be of high interest.

In relation to the K-edge XMCD experiments\cite{Ishimatsu-paramag-Co,Torchio-Co-120GPa} it is worth mentioning that a 90-degree SS state does not have any magnetic dichroism, because all components of the magnetisation are compensated on a macroscopic scale.
On the other hand, an application of the external magnetic field is necessary in these experiments.
Large fields, in principle, could lead to the bending of the SS, thus resulting in a different cone angle of the spiral. 
Our results suggest that, even if $\Theta$ reaches 30 degrees, these SS states will be more favoured than the FM one (see Fig.~\ref{spirals}).
However, such a state would already be characterised by a non-zero XMCD signal, since the total $M_S$ will be finite.

At this point it is important to discuss the relativistic effects, so far neglected in our study.
Most importantly, spin orbit coupling gives rise to the magneto-crystalline anisotropy energy (MAE), which defines a preferable direction (or plane) for spins to point to.
At ambient conditions the MAE of hcp Co is 4.77 $\mu$Ry per atom.\cite{3dMe-MAE}
This value is one order of magnitude smaller than the maximal difference between the total energy of the SS and the FM state.
Moreover, the most favourable SS configuration was found above 100 GPa. 
At such large compressions the orbital moment is strongly suppressed, since it attenuates faster than the spin moment.\cite{Torchio-ni-prl,hcpCo-hydrostat}
The MAE is supposed to follow the orbital moment, since their relation is known.\cite{pbruno-MAE89}
Thus, we are confident that our conclusions remain valid in the case of finite spin-orbit coupling.

In summary, we have shown that a sequence of ETTs happen in hcp Co upon compression.
The most important transition, occurring at 80 GPa, is suggested to be the reason for the number of anomalies in the magnetic, elastic and vibrational properties, confirmed both experimentally and theoretically.
In the high-pressure regime, where the magnetic moment of Co is reduced to $\approx1.3\mu_B$, the ETT is demonstrated to lead to the stabilisation of non-collinear magnetic states.

The authors acknowledge the computational resources provided by the Swedish National Infrastructure for Computing (SNIC) and Uppsala Multidisciplinary Center for Advanced Computational Science (UPPMAX).
Support from KAW foundation, VR and eSSENCE are acknowledged.


%

\pagebreak

\section{Supplemental Material}
\subsection{Band structure under pressure}

The Fermi surface (FS) changes, presented in the main text of the article, clearly show the presence of the series of electronic topological transitions (ETTs)  in hcp Co upon compression.
Another way of observation of the ETT is an analysis of the band structure.
In Fig.~1 we show the $k$-resolved spectral function for different pressures.

\begin{figure*}[!h]  
\includegraphics[angle=0,width=180mm]{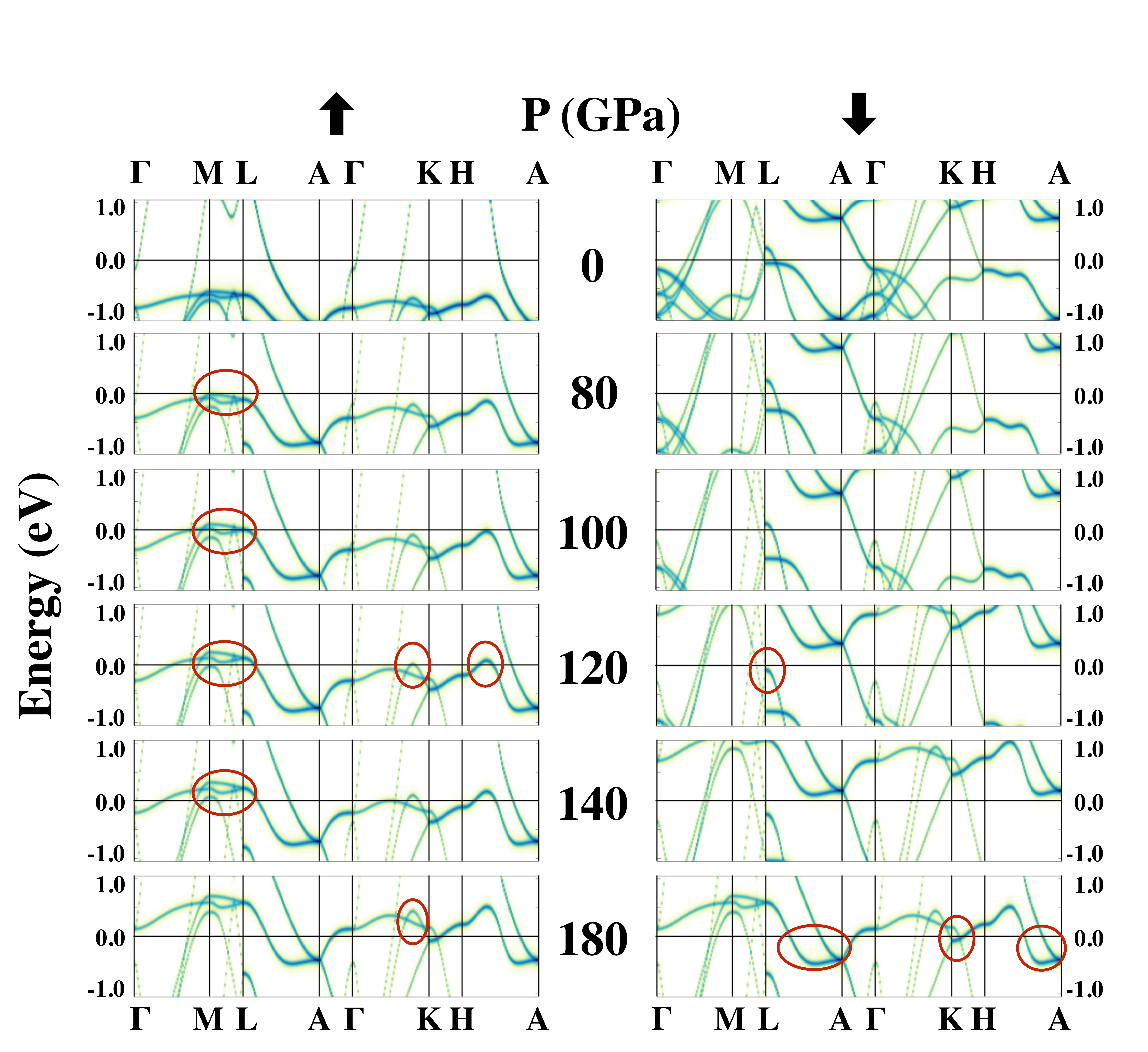}
\caption{(Color on-line) Computed $\vec k$-resolved spectral functions of the ferromagnetic hcp Co at different pressures. 
Spin-up bands are shown on the left side of the plot, while spin-down states are on the right one.
The results were obtained with the \texttt{RSPt} code. Red circles underline the bands, which cross the Fermi surface at different compressions. The Fermi level is set to zero.}
\label{bands}
\end{figure*}

An inspection of Fig.~1 reveals that the most of the ETT's occur for spin-up bands.
Above 80 GPa there are several transitions particularly along M-L direction in the Brillouin Zone, which is parallel to the $z$ axis.
These transitions create a larger overlap between majority and minority states. 
As a result, a particular family of spin spirals, which is characterised by $\vec q$ parallel to $z$, becomes preferable at high pressure.

\subsection{ETT and other properties}
In this section we provide a direct evidence that the ETT's in compressed hcp Co give rise to the anomalies in $c/a$ ratio and magnetic moment ($M_S$).
In Fig.~2 the pressure dependence of both quantities are presented along with the corresponding values of the density of states (DOS) at the Fermi level.
The latter quantity varies highly non-monotonously under pressure, because of the sequence of the ETT's happening upon compression.
However, the most remarkable changes in the DOS($E_F$) and its pressure derivative occur around 80 GPa.
This is the main reason why this particular ETT affects so drastically the properties of the material.

The trends of the magnetic moment and $c/a$ ratio are very similar to the ones obtained by Antonangeli \textit{et al.} (Ref.~\onlinecite{antonangeli-Co-APL2008}).
The differences in the reported results are more pronounced for the $c/a$ ratio and clearly originate from the different basis sets used in the calculations.
The authors of Ref.~\onlinecite{antonangeli-Co-APL2008} employed full-potential linearized augmented plane wave method (FP-LAPW), while we used projector augmented wave method (PAW).
Moreover, $c/a$ ratio is a parameter, which is hard to converge. Its final values might depend on the method for its extraction. 
We used the stress tensor calculations instead of the parabolic fit of total energy as was done in Ref.~\onlinecite{antonangeli-Co-APL2008}.
It is also worth emphasising that we have used a much denser $k$-point grid containing 45$\times$45$\times$27 points, as compared with Antonangeli \textit{et al.} who used 16$\times$16$\times$8 points.

In spite of all these details, both reported trends are qualitatively the same and show anomalies around 75-80 GPa.
Fig.~2 re-considers these results in the view of an ETT, which we observed.

\begin{figure}[!h]  
\includegraphics[angle=0,width=100mm]{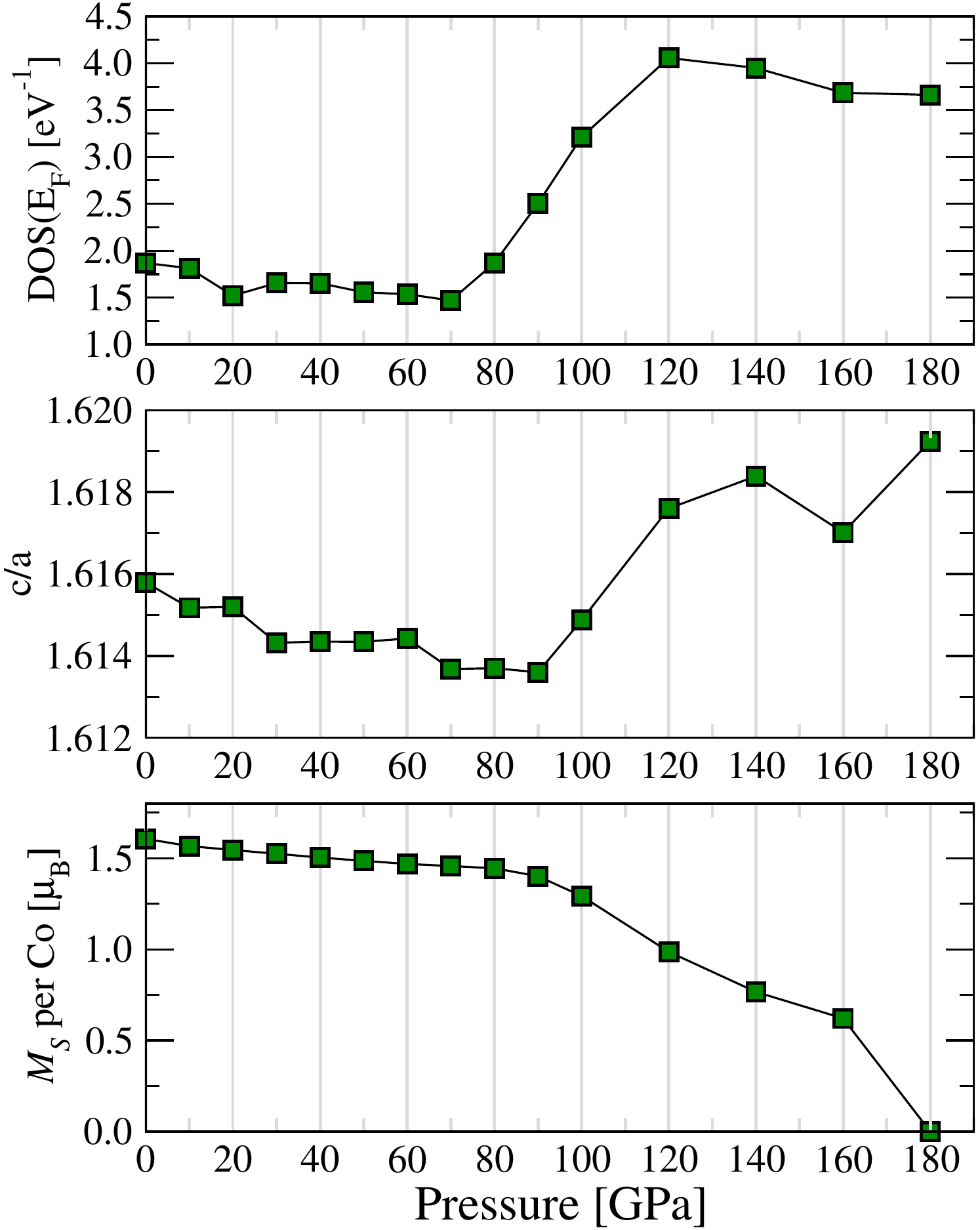}
 \caption{(Color on-line) Pressure dependence of several ground state properties. Upper panel: DOS at the Fermi level. Middle panel: optimised $c/a$ ratio. Lower panel: Magnetic moment ($M_S$) per Co atom.}
\label{moms}
\end{figure}

\end{document}